\documentclass[pra,twocolumn,showpacs,floatfix,a4paper,superscriptaddress]{revtex4}
\usepackage{bm,color,graphicx,amsmath,txfonts}
\usepackage{hyperref}

\newcommand{\rp}[1]{(\ref{#1})}

\newcommand{\abs}[1]{\left|{#1}\right|}

\newcommand{\da}{^\dagger}

\newcommand{\pt}[1]{\left( #1 \right)}
\newcommand{\pq}[1]{\left[ #1 \right]}

\newcommand{\nn}{{\nonumber}}

\newcommand{\GG}{{\cal G}}

\begin{document}

\title{Einstein-Podolsky-Rosen steering and Bell nonlocality of two macroscopic mechanical oscillators \\
in optomechanical systems}

\author{Jie Li}
\affiliation{Department of Physics, Zhejiang University, Hangzhou 310027, China}
\affiliation{Texas A{\rm \&}M University, College Station, Texas 77843, USA}
\author{Shi-Yao Zhu}
\affiliation{Department of Physics, Zhejiang University, Hangzhou 310027, China}

\begin{abstract}
We investigate under which conditions quantum nonlocal manifestations as Einstein-Podolsky-Rosen steering or Bell nonlocality can manifest themselves even at the macroscopic level of two mechanical resonators in optomechanical systems. We adopt the powerful scheme of reservoir engineering, implemented by driving a cavity mode with a properly chosen two-tone field, to prepare two mechanical oscillators into an entangled state. We show that large and robust (both one-way and two-way) steering could be achieved in the steady state with realistic parameters. We analyze the mechanism of the asymmetric nature of steering in our system of two-mode Gaussian state. However, unlike steering, Bell nonlocality is present under much more stringent conditions. We consider two types of measurements, displaced parity and on-off detection, respectively. We show that for both the measurements Bell violation requires very low environmental temperature. For the parity detection, large Bell violation is observed only in the transient state when the mechanical modes decouple from the optical mode and with extremely small cavity losses and mechanical damping. Whereas for the on-off detection, moderate Bell violation is found in the steady state and robust against cavity losses and mechanical damping. Although Bell violation with the parity detection seems extremely challenging to be experimentally demonstrated, the conditions required for violating Bell inequalities with the on-off detection are much less demanding.

\end{abstract}

\date{\today}
\maketitle

\section{Introduction}

In 1964, Bell proved that no local realistic theory can completely describe the predictions of quantum mechanics, which is known as Bell's theorem~\cite{Bell}. He showed the limitations imposed by local realism in the form of inequalities. The violation of Bell inequalities implies that the correlations between the outcomes of measurements made upon composite systems cannot be explained by local realistic theories, or simply, nonlocal correlations are present within the system. There is another form of nonlocality, namely, Einstein-Podolsky-Rosen (EPR) steering. It was first pointed out in 1935 by EPR~\cite{EPR} and later discussed by Schr\"odinger~\cite{Schrodinger}. It describes a phenomenon that two distant parties share an entangled state and one party, by measuring its subsystem, can remotely change the state of the other party's subsystem. Like Bell nonlocality \cite{Bell,BellRMP}, EPR steering is demonstrated by the violation of steering inequalities~\cite{Wiseman}. However, unlike Bell nonlocality and entanglement~\cite{enRMP}, steering has a fundamental asymmetric property in the sense that in a steering test the two parties play a different role: there exist entangled states which are only steerable from one party to the other party (i.e., one-way steering), but not vice versa. Such an asymmetric feature has important applications for the task of one-sided device-independent quantum key distribution~\cite{Wiseman12,Walk}. Typically, steering is considered as a form of quantum correlation that lies in between entanglement and Bell nonlocality: nonlocality implies two-way steerability, while one-way steerability implies entanglement, and the converse relations do not hold~\cite{relations}.

Nonlocality has recently been demonstrated by the violation of Bell inequalities free of both locality and detection loopholes in photonic systems~\cite{Bellexp}, and by the violation of steering inequalities in a number of experiments~\cite{EPRexp}. However, these demonstrations have been done only in microscopic systems. Nonlocality has not yet been observed in mesoscopic or macroscopic systems, e.g., between two massive mechanical oscillators. The studies of the possibility of observing quantum correlations shared by two macroscopic objects are of fundamental importance since they are related to the research of quantum-to-classical transition~\cite{Zurek}, wave-function collapse theories~\cite{Bassi,Jie}, macroscopic quantum mechanics~\cite{Chen,macroRMP}, and so on.

In this paper, we study the nonlocal properties, EPR steering and Bell nonlocality, of two macroscopic mechanical resonators (MRs) in optomechanical systems. Optomechanics, addressing the coupling between optical and mechanical degrees of freedom via radiation pressure, provides a promising platform to observe quantum effects in mechanical systems~\cite{OMRMP,ccool,ssqueez,Simon17,mechEn}. In order to test steering and Bell inequalities in mechanical systems, one should first prepare two MRs into an entangled state. Many schemes have been proposed for the generation of entanglement between two MRs in optomechanical systems. They exploit, for example, radiation pressure~\cite{PRL02,jopa,genesNJP,hartmann}, transfer of entanglement~\cite{Peng03,Reid17} and squeezing~\cite{EPL} from optical fields, conditional measurements on light modes~\cite{entswap,bjorke,mehdi1,woolley,mehdi2,Savona}, and reservoir engineering implemented by a properly chosen two-tone driving~\cite{Clerk,WoolleyClerk,Buchmann,JieNJP,JieCF}. In the present work, we adopt the schemes of Refs.~\cite{JieNJP,JieCF} which are able to generate, either dynamically or in the steady state, large entanglement between two MRs. The scheme of Ref.~\cite{JieCF} is the improved version of Ref.~\cite{JieNJP} by including a coherent feedback loop, which reduces the effective cavity decay rate resulting in a remarkable enhancement of the mechanical entanglement.

We first study the EPR steering of the MRs and find that large and robust steering could be generated in the steady state with realistic parameters. We show optimal working conditions for obtaining large steering and analyze the mechanism for the asymmetric nature of steering in our system. We find that such an asymmetric nature is due to the difference of quantum fluctuations of the two mechanical modes and the asymmetry disappears for equal fluctuations of the two modes. Furthermore, we discuss the hierarchical relationships of entanglement, one-way and two-way steerings in our system of two-mode Gaussian states. We then analyze if and when the entangled state violates Bell inequalities constructed in terms of the correlation functions of two different observables. Specifically, we consider the observables corresponding to displaced parity and on-off measurements, respectively. We find that for displaced parity measurement large Bell violation is present only in the transient regime and it requires very low environmental temperature and extremely small cavity losses and mechanical damping. On the contrary, for displaced on-off measurement, moderate Bell violation is found in the steady state even in the case of significant losses. This allows for the possibility of testing Bell nonlocality of two macroscopic MRs in the near future.

The paper is organized as follows. In Sec.~\ref{system}, we introduce our system that is used to prepare two MRs into an entangled state. We then study the EPR steering of the MRs in Sec.~\ref{SecSteering}. We discuss the relationships between entanglement and one-way, two-way steerings in our specific model. In Sec.~\ref{SecBell}, we test the Bell nonlocality in phase space with displaced parity and on-off measurements, respectively. We show the parameter regime within which the Bell inequality is violated. Finally, we make our conclusions in Sec.~\ref{concl}.

\section{The system}
\label{system}

We consider two MRs with different frequencies $\omega_1$ and $\omega_2$ within an optical Fabry-P\'erot cavity. The two MRs interact via the usual optomechanical interaction with a cavity mode of frequency $\omega_c$, which is bichromatically driven at the two frequencies $\omega_{L1}=\omega_0+\omega_1$ and $\omega_{L2}=\omega_0-\omega_2$, where $\omega_0$ is the reference frequency and is slightly detuned from the cavity resonance by $\Delta_0=\omega_c-\omega_0$. In other words, the cavity mode is simultaneously driven close to the blue sideband associated with the MR of frequency $\omega_1$ and close to the red sideband associated with the MR of frequency $\omega_2$. The Hamiltonian of the system in the reference frame rotating at the frequency $\omega_0$ reads~\cite{JieNJP}
\begin{equation}
\begin{split}
\hat{H}&=\hbar\Delta_0  \hat{a}^\dagger \hat{a} + \hbar\sum_{j=1}^2\omega_j \hat{b}^\dagger_j \hat{b}_j + \hbar \sum_{j=1}^2 g_j  \hat{a}^\dagger \hat{a} \left(\hat{b}_j+\hat b_j^\dagger\right)  \\
&+\hbar\pq{\pt{E_1 e^{-i\omega_1t}+E_2 e^{i\omega_2t}}\hat{a}^\dagger +{\rm H.c.}},
\end{split}
\label{haml}
\end{equation}
where $\hat{a}$ ($\hat{b}_{1,2}$) is the annihilation operator of the cavity mode (mechanical modes), $g_j$ is the bare optomechanical coupling associated with the $j$th MR, and $E_j=\! \sqrt{2 P_j \kappa_1/\hbar \omega_{Lj}}$, where $P_j$ is the power of the driving field and $\kappa_1$ and $\kappa_2$ are, respectively, the cavity decay rates due to the transmission through the two cavity mirrors.

The system dynamics can be efficiently studied by linearizing the optomechanical interaction in the limit of strong
driving fields. The relevant degrees of freedom for the linearized dynamics are the fluctuations of the cavity field and the mechanical modes about their respective average values. Unlike the standard approach adopted in the analysis of optomechanical systems~\cite{OMRMP}, here the average fields are time dependent as a result of the bichromatic driving field. Nevertheless, approximated, time independent equations for the system dynamics can be derived by focusing only on the dominant resonant processes, and the non-resonant processes can be safely neglected if the following conditions are fulfilled~\cite{JieNJP}
\begin{eqnarray}\label{cond01}
\abs{g_j E_j/\omega_j}  ,\,\, \kappa_{1,2}  \ll \omega_{1,2}, \, \abs{\omega_1-\omega_2} .
\label{condQLEs}
\end{eqnarray}
Eq.~\rp{cond01} implies significantly different mechanical frequencies in order to suppress unwanted optomechanical processes~\cite{JieNJP}, and sets stringent constrains due to the relatively small mechanical frequencies that typically characterize massive resonators, of which the nonlocal properties are what we are interested in. In practice, the restriction on the optomechanical couplings $\abs{g_j E_j/\omega_j}  \ll \omega_{1,2}, \, \abs{\omega_1-\omega_2}$ can be easily satisfied by lowering the power of the driving field, while the condition on the cavity decay rates $\kappa_{1,2}  \ll \omega_{1,2}, \, \abs{\omega_1-\omega_2}$ is more difficult to be met. However, as shown in Ref.~\cite{JieCF}, by including a proper coherent feedback loop, which would reduce the effective cavity decay rate, the above condition can be largely relaxed. Furthermore, the entanglement of the MRs can be enhanced due to an enhanced cooperativity. This is important since, in general, only when the entanglement is strong enough one-/two-way steering and Bell violation appear.

\begin{figure*}[t]
\hskip-0.2cm\includegraphics[width=16cm]{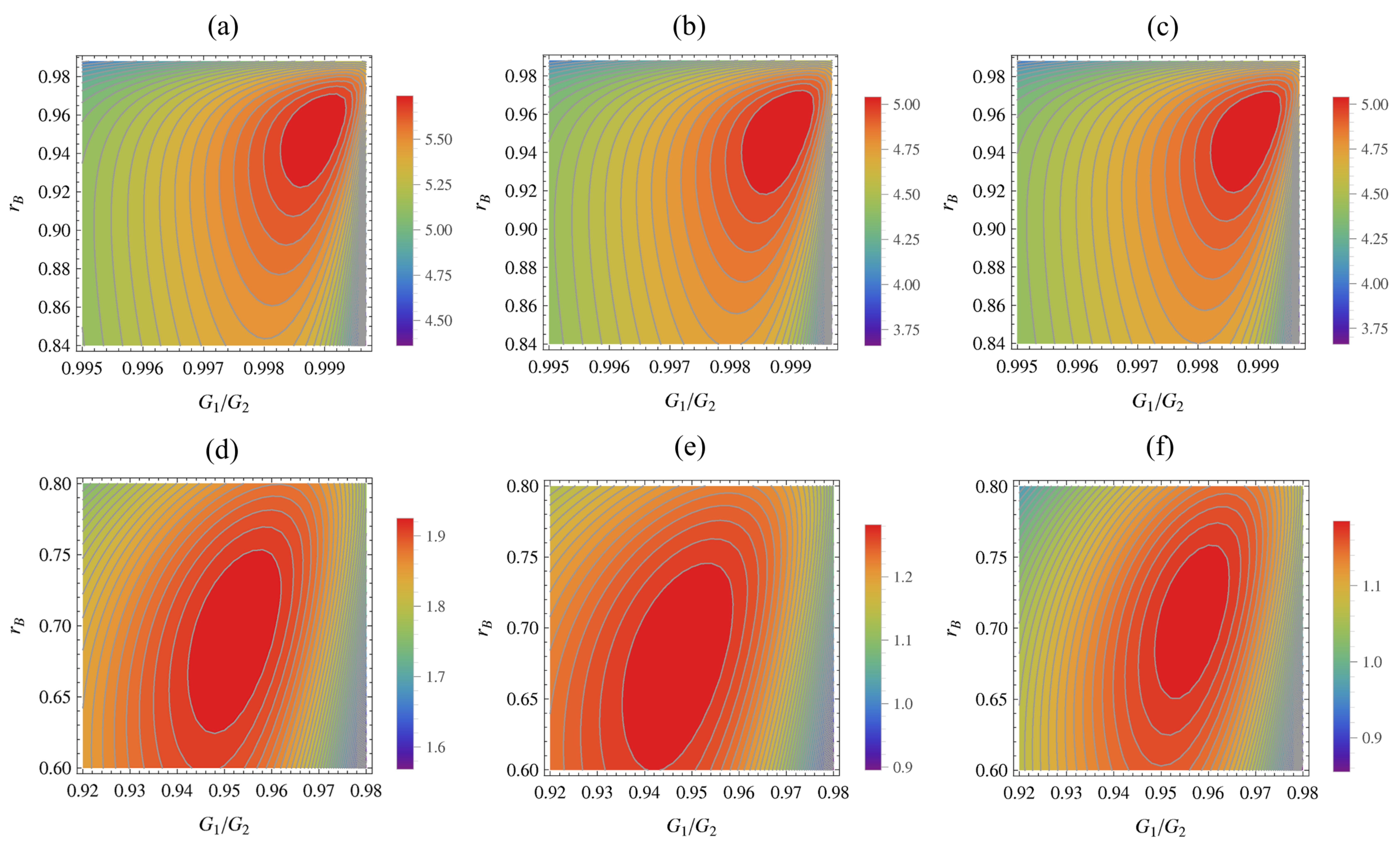}
\caption{Contour plot of the steady-state entanglement (logarithmic negativity) $E_N$ (left), steering $\GG^{1\to 2}$ (middle) and $\GG^{2\to 1}$ (right) as a function of $G_1/G_2$ and $r_B$ with (a)-(c) $\bar n_1=\bar n_2=0$ and (d)-(f) $\bar n_1=2\bar n_2=200$. We have taken $\gamma_1=\gamma_2=10$ Hz, $G_2=2\kappa_1=2\kappa_2=10^5$ Hz, $\Delta=0$ and $\theta=0$. For these parameters, the effective cavity decay rate is simply $\tilde\kappa=2\kappa_1 (1-r_B)$. }
\label{fig1}
\end{figure*}

With the conditions~\rp{cond01} fulfilled, the dynamics of the system with coherent feedback can be described by the following set of quantum Langevin equations, which in the interaction picture with respect to $\hat{H}_0=\hbar\sum_{j=1}^2\omega_j \hat{b}^\dagger_j \hat{b}_j$ are given by~\cite{JieCF}
\begin{eqnarray}
 && \delta \dot{\hat{a}}=-(\tilde\kappa+i\,\tilde\Delta) \delta \hat{a}-i G_1 \delta \hat{b}_1^\dagger -i G_2 \delta \hat{b}_2 +\!\sqrt{2\,\tilde\kappa}\, \hat A^{\rm in},  \label{a1st} \\
 &&\delta\dot{\hat{b}}_1 =-\frac{\gamma_1}{2} \delta \hat{b}_1  - i G_1 \delta \hat{a}^{\dagger}+\!\sqrt{\gamma_1} \hat{b}_1^{\rm in},  \label{b1st} \\
&&\delta\dot{\hat{b}}_2 =-\frac{\gamma_2}{2} \delta \hat{b}_2  - i G_2^* \delta \hat{a}+\!\sqrt{\gamma_2} \hat{b}_2^{\rm in}, \label{b2st}
\end{eqnarray}
where $\tilde\kappa$ and $\tilde\Delta$ are the effective cavity decay rate and detuning modified by the feedback, given by~\cite{JieCF}
\begin{eqnarray}\label{tildekappaDelta}
\tilde\kappa&=&\kappa_1+\kappa_2-2\sqrt{\kappa_1\,\kappa_2}\ r_B \, \cos\theta,
\nn\\
\tilde\Delta&=&\Delta-2\sqrt{\kappa_1\,\kappa_2}\ r_B \, \sin\theta,
\end{eqnarray}
where the detuning $\Delta$ includes the frequency shift due to the optomechanical interaction~\cite{JieNJP}, and $r_B$ and $\theta$ are two parameters related to the feedback loop~\cite{JieCF}: $r_B$ is the reflection coefficient of the controllable beamsplitter and $\theta$ is the phase shift of the light in the feedback loop. From Eq.~\eqref{tildekappaDelta}, we see that the cavity decay rate can be significantly reduced when the cavity is symmetric with $\kappa_1\,{=}\,\kappa_2$, the reflectivity approaches unity, $r_B \, {\to}\, 1$, and the phase shift $\theta\,{=}\, 2m\pi$ ($m\,{=}\,0,1,2,...$). For these values of $\theta$, the detuning remains unchanged $\tilde\Delta=\Delta$. $\gamma_1$ and $\gamma_2$ are the damping rates of the two mechanical modes, respectively. $G_1\,{=}\,g_1 E_1/(\omega_1{-}\tilde\Delta {+} i \tilde\kappa)$ and $G_2\,{=}\,g_2 E_2/(-\omega_2{-}\tilde\Delta {+} i \tilde\kappa)$ are the effective optomechanical couplings. $\hat A^{\rm in}=\Big[ (\! \sqrt{\kappa_2}\,{-}\!\sqrt{\kappa_1} e^{i \theta}\, r_B) \,\hat{a}_2^{\rm in}+\!\!\sqrt{\kappa_1 (1-r_B^2)} \,\hat{a}_1^{\rm in} \Big] /\sqrt{\tilde\kappa}$ is the new input noise operator modified by the feedback and satisfies the correlation function $\langle\hat{A}^{\rm in} (t) \hat{A}^{\rm in}{}^\dag (t')\rangle{=}\delta (t{-}t')$. $\hat{a}_1^{\rm in}$ and $\hat{a}_2^{\rm in}$, instead, denote the original input noises without feedback entering the two cavity mirrors~\cite{JieCF}, and their nonzero correlation functions are $\langle\hat{a}_i^{\rm in}(t)\, \hat{a}_i^{\rm in}(t')\da \rangle=\delta(t{-}t')$. $\hat{b}_j^{\rm in}$ describes the noise of the $j$th MR and its correlation functions are $\langle\hat{b}_j^{\rm in}(t)\, \hat{b}_j^{\rm in}(t')\da \rangle \,{=}\, (\bar n_j{+}1) \, \delta(t{-}t')$ and $\langle\hat{b}_j^{\rm in}(t)\da\, \hat{b}_j^{\rm in}(t') \rangle \, {=}\, \bar n_j\delta(t{-}t')$, with $\bar{n}_j=\big[\!\exp (\hbar \omega_j/k_B T )-1 \big]^{-1}$ the mean thermal phonon number which is assumed to stay at the same environmental temperature $T$.

Since the dynamics of the system is linearized and all noises are Gaussian, the dynamical map of the system preserves the Gaussian nature of any input state. In this situation, the system state is completely characterized by the first and second moments of the quadrature operators. Since we are interested here in the correlation properties of the two MRs, the first moments will not be relevant and we thus discard them. The second moments can be arranged in the form of a covariance matrix (CM) $V(t)$ with its entries defined as $V_{ij}=\frac{1}{2}\langle \{ \hat{u}_i(t), \hat{u}_j(t) \} \rangle$, where $\{ \cdot , \cdot \}$ denotes an anticommutator and $\hat{u}$ is the vector of quadrature fluctuation operators of the two mechanical modes, i.e., $\hat{u}(t)=\big (\delta \hat{q}_1(t),  \delta\hat{p}_1(t),  \delta\hat{q}_2(t),  \delta\hat{p}_2(t) \big )$, with $\delta \hat{q}_j{=}(\delta \hat{b}_j{+}\delta \hat{b}_j^{\da})/\!\sqrt{2}$, $\delta \hat{p}_j{=} i (\delta \hat{b}_j^{\da}{-}\delta \hat{b}_j)/\!\sqrt{2}$ ($j{=}1,2$). The CM $V(t)$ at any time $t$ can be obtained following the method provided in the Appendix of Ref.~\cite{JieCF}. Here we will not reiterate it but present the results directly in the next sections.

\begin{figure*}
\hskip0.55cm{\bf (a)}\hskip5.3cm{\bf (b)}\hskip5.4cm{\bf (c)}\\
\hskip-0.1cm\includegraphics[width=17cm]{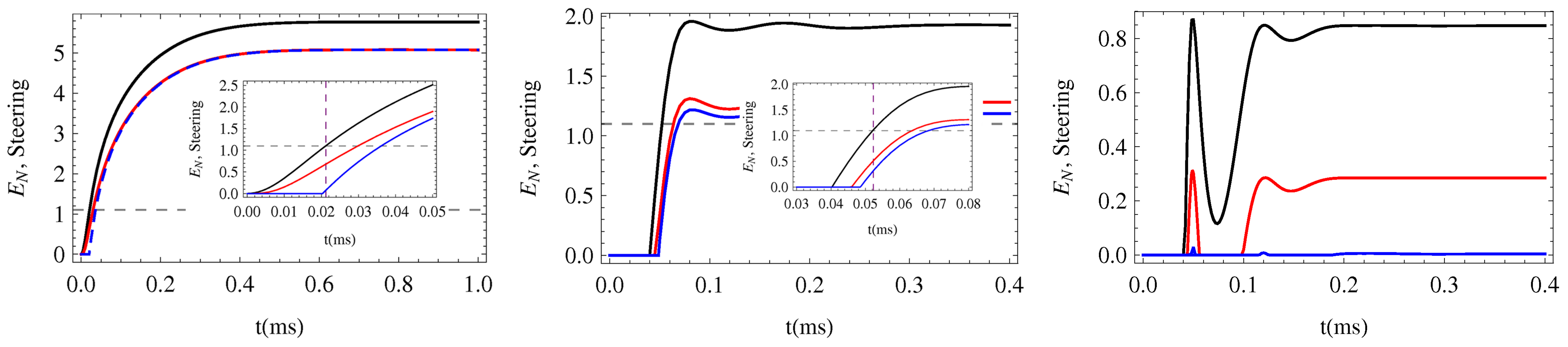}  
\caption{Time evolution of steering $\GG^{1\to 2}$ (red lines), $\GG^{2\to 1}$ (blue lines) and entanglement $E_N$ (black lines) for (a) $r_B=0.95$, $G_1=0.999 G_2$ and $\bar n_1=\bar n_2=0$, (b) $r_B=0.7$, $G_1=0.953 G_2$ and $\bar n_1=2\bar n_2=200$, (c) $r_B=0.5$, $G_1=0.869 G_2$ and $\bar n_1=2\bar n_2=1000$. The gray dashed lines denote $E_N={\rm ln} 3$, above which both $\GG^{1\to 2}$ and $\GG^{2\to 1}$ are nonzero, i.e., two-way steerable. This is clearly seen in the insets. The other parameters are as in Fig.~\ref{fig1}. }
\label{fig2}
\end{figure*}

\section{EPR Steering at steady state}
\label{SecSteering}

In this section, we study the EPR steering of the MRs. We adopt the measure of Ref.~\cite{Adesso} which is defined for arbitrary bipartite Gaussian states of continuous variable (CV) systems under Gaussian measurements. For the simplest case of two-mode Gaussian states, it assumes the following simple form
\begin{equation}
\GG^{1\to 2}(V)={\rm max}\{0, S(2V_1)-S(2V) \},
\label{steerDef}
\end{equation}
and for defining $\GG^{2\to 1}(V)$ by replacing $S(2V_1)$ with $S(2V_2)$. $V_1$ and $V_2$ are $2 \times 2$ CMs corresponding to the reduced states of subsystems of MR 1 and 2, respectively. $S$ is the R\'enyi-2 entropy, which for a Gaussian state with CM $\sigma$ is given by $S(\sigma)=\frac{1}{2} {\rm ln} ({\rm det} \sigma)$~\cite{Adesso2}. Note that there is a difference of a factor of 2 in $S(\cdot)$ between the definition~\eqref{steerDef} and Eq.\!\! (5) of Ref.~\cite{Adesso} due to their different definitions of CM. A nonzero $\GG^{1\to 2}(V)$ ($\GG^{2\to 1}(V)$) denotes that the state described by CM $V$ is steerable from MR 1 (2) to 2 (1) by applying Gaussian measurements on MR 1 (2), and its value quantifies the amount by which the steering inequality is violated~\cite{Adesso}. $\GG^{1\to 2}$ and $\GG^{2\to 1}$ are generally different quantities and they are equal when the CMs of two subsystems are identical. If a state has both nonzero $\GG^{1\to 2}$ and $\GG^{2\to 1}$ we call the state two-way steerable, otherwise we call the state either only one-way steerable or nonsteerable.

Steering has been investigated in optomechanical systems mainly focusing on the quantum correlations between mechanical and optical degrees of freedom~\cite{He1,He2,Tan}. In Ref.~\cite{Tan}, steering has been studied between two MRs of which the entangled state is prepared via entanglement swapping. There two identical MRs have been considered which results in the absence of the asymmetry of steering. Instead, we adopt a different entanglement generation scheme and the unequal couplings $G_2>G_1$ required for the system stability yield generally different CMs $V_{1}$ and $V_{2}$ which, according to the definition~\eqref{steerDef}, lead to generally different steerings $\GG^{1\to 2} \, {\ne} \, \GG^{2\to 1}$. In our system, the entangled state of the MRs is a two-mode squeezed state and the reduced state of each MR is a purely thermal state~\cite{JieNJP} with CM $V_1{=}{\rm diag}(a,a)$ for MR 1 and $V_2{=}{\rm diag}(b,b)$ for MR 2, where $a$ and $b$ denote the variance of the quadrature fluctuations of MR 1 and 2, respectively, i.e., $a \equiv \langle \delta \hat{q}_1^2 \rangle=\langle \delta \hat{p}_1^2 \rangle$ and $b \equiv \langle \delta \hat{q}_2^2 \rangle=\langle \delta \hat{p}_2^2 \rangle$. In fact, $G_2>G_1$ implies that MR 2 that is driven on the red sideband (corresponding to the process of removing mechanical excitations) is more strongly coupled to the light with respect to MR 1 that is driven on the blue sideband (corresponding to the process of adding mechanical excitations). In our system, the fluctuation of the quadratures of MR 2 is always smaller than that of MR 1 in the steady state, i.e., $a>b$, implying that $S(2V_1) > S(2V_2)$, which, according to the definition \eqref{steerDef}, means $\GG^{1\to 2}>\GG^{2\to 1}$. Physically this could be interpreted that, for a two-mode Gaussian state (under Gaussian measurements), the mode with a lower excitation number, or fluctuation, is easier to be steered by the other mode with a higher excitation number. Such a feature has been demonstrated by the results of Figs.~\ref{fig1} and \ref{fig2}.

In Fig.~\ref{fig1} we show two different direction steerings $\GG^{1\to 2}$ and $\GG^{2\to 1}$ of the two MRs as a function of two key parameters $G_1/G_2$ and $r_B$ at different temperatures, and compare them with the entanglement $E_N$ quantified by logarithmic negativity~\cite{logNeg} in consideration of their similar definitions for Gaussian states~\cite{Adesso}. The mechanism for the presence of optimal values of $G_1/G_2$ and $r_B$ has been expounded in Ref.~\cite{JieCF}. For simplicity, we have assumed $\omega_2\,{=}\,2\omega_1$ in all the figures throughout the paper. For the ideal case of zero temperature $\bar n_{1,2}=0$, the optimal values of $G_1/G_2$ and $r_B$ are almost the same for $\GG^{1\to 2}$, $\GG^{2\to 1}$ and $E_N$, and the two steerings are almost equal. This is because in this case $G_1$ and $G_2$ are so close that the CMs $V_1$ and $V_2$ have little difference leading to the fact that $\GG^{1\to 2} \approx \GG^{2\to 1}$. As the temperature rises, the optimal couplings $G_1$ and $G_2$ will have a larger difference (or, a lower ratio of $G_1/G_2$ for fixed $G_2$)     
in order to generate large entanglement~\cite{JieCF}. A larger difference of $G_1$ and $G_2$ will eventually lead to a larger difference of $\GG^{1\to 2}$ and $\GG^{2\to 1}$, as explained in the previous paragraph and shown in Figs.~\ref{fig1} and \ref{fig2}. In Fig.~\ref{fig1} (d)-(f), the optimal values of $G_1/G_2$ and $r_B$ for $\GG^{1\to 2}$, $\GG^{2\to 1}$ and $E_N$ are no longer overlapped and start to separate to what extent depending on the temperature.

Fig.~\ref{fig2} shows the time evolution of $\GG^{1\to 2}$, $\GG^{2\to 1}$ and $E_N$, which lead to a steady state, for different temperatures with $\bar n_1=2\bar n_2=0$, 200, 1000. It is evident that the two steerings and entanglement show similar behaviors due to their similar definitions for Gaussian states. We have verified in our specific model (see the insets of Fig.~\ref{fig2} (a)-(b)) that the state with logarithmic negativity $E_N \, {>} \, {\rm ln} 3 \, {\approx} \, 1.1$ is of two-way steerability. This is valid for any two-mode Gaussian state under Gaussian measurements~\cite{He3,Adesso}. Fig.~\ref{fig2} (c) shows that one-way steering is a type of quantum correlations that is stronger than entanglement and only when the entanglement is strong enough it occurs.

\begin{figure*}[t]
\hskip0.6cm{\bf (a)}\hskip7.2cm{\bf (b)}\\
\hskip-0.75cm\includegraphics[width=15.3cm]{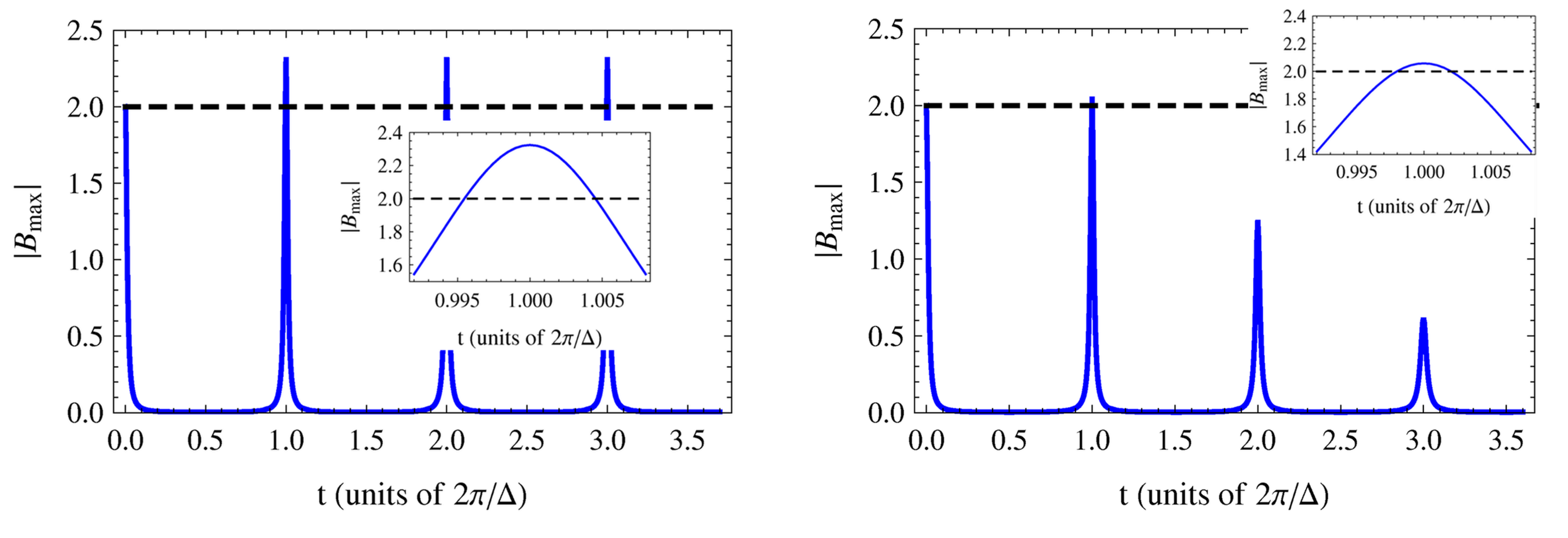}   \\
\caption{Time evolution of $|{\cal B}_{\rm max}|$ in the dynamical case of $G_1=G_2=10^5$ Hz. The evolution starts from an initial separable state of the cavity mode in the vacuum state and each MR in its thermal state with mean thermal phonon number $\bar n_{1,2}$. (a) $\gamma_1=\gamma_2=0$; (b) $\gamma_1=\gamma_2=0.3$ mHz. The dashed lines denote $|{\cal B}_{\rm max}|=2$ and the insets show the Bell violation about $t=2\pi/\Delta$. The other parameters are $\Delta=10^4$ Hz, $\tilde \kappa=0$, $\bar n_1=\bar n_2=0$ and $\theta=0$. }
\label{fig3}
\end{figure*}

\begin{figure*}
\hskip-0.37cm{\bf (a)}\hskip7.7cm{\bf (b)}\\
\hskip-0.28cm\includegraphics[width=16cm]{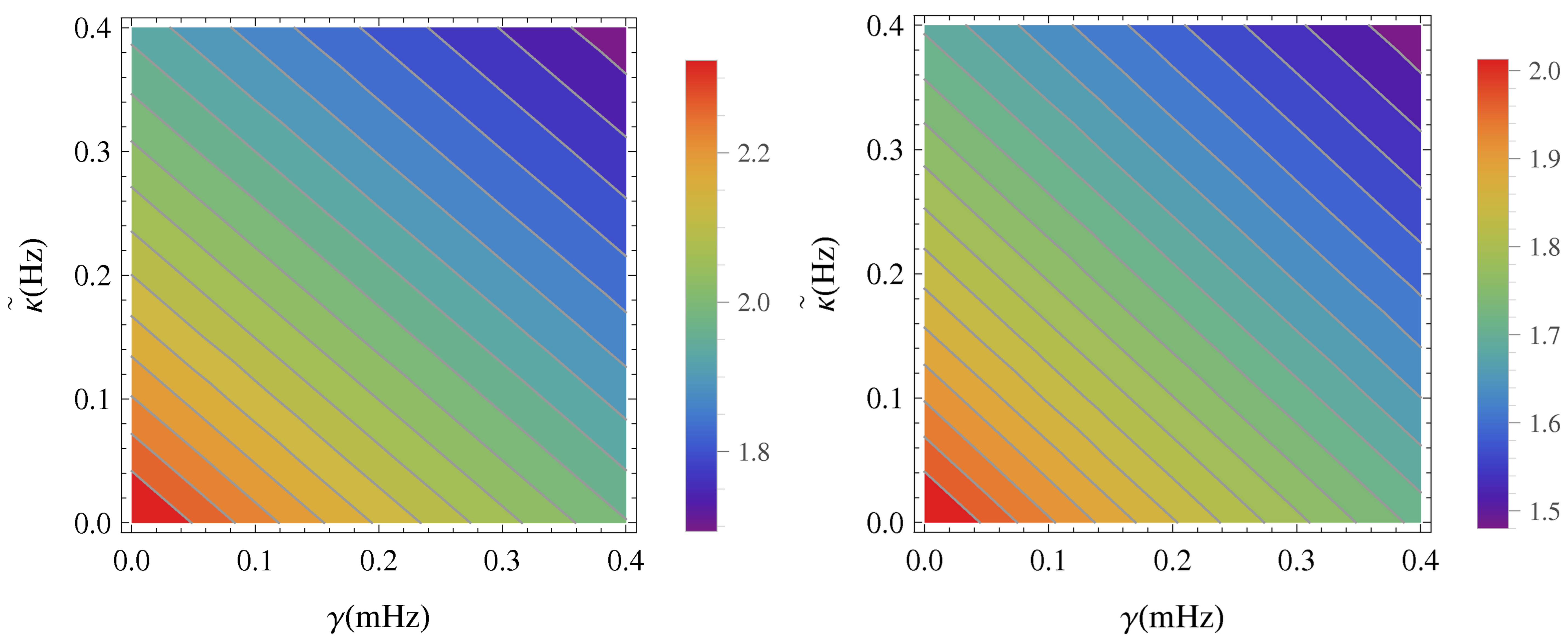}   \\
\caption{Contour plot of $|{\cal B}_{\rm max}|$ versus $\gamma_1=\gamma_2 \equiv \gamma$ and $\tilde \kappa$ in the dynamical case of $G_1=G_2=10^5$ Hz. (a) $\bar n_1=\bar n_2=0$; (b) $\bar n_1=2\bar n_2=0.05$. We take the optimal time $t=2\pi/\Delta$, $\Delta=10^4$ Hz and $\theta=0$. }
\label{fig4}
\end{figure*}

\section{Testing Bell nonlocality in phase space}
\label{SecBell}

Having observed strong one-way and two-way steerings between the MRs in the steady state, in this section we devote ourselves to the study of another type of nonlocality, namely Bell nonlocality, demonstrated by the violation of a proper Bell inequality. Proposals to test Bell inequalities in an optomechanical system have been put forward~\cite{Sangouard,Klemens}, focusing on the correlations between the optical and mechanical degrees of freedom. Here, instead, we study the nonlocal correaltions between two MRs. Typically, Bell nonlocality is considered as the strongest quantum correlation that is stronger than any other types such as two-way, one-way steering and entanglement~\cite{relations}. This is indeed the case in our system, as we will show later, the presence of Bell nonlocality requires much more stringent conditions than those for steering and entanglement. We shall test Bell inequalities based on two observables corresponding to displaced parity and on-off detection, respectively. Using the fact that the mean value of these two measurements are proportional to the quasiprobability functions, this allows one to perform Bell tests in phase space~\cite{BW98,BW99,Lee}.

\subsection{Bell violation with parity detection}
\label{Bell-W}

Nonlocality of CV systems can be tested in phase space by making the displaced parity measurement on each mode~\cite{BW98,BW99,Lee}. Such a phase-space approach is based on the fact that the expectation value of the displaced parity operator is linked to the Wigner function~\cite{BW98,BW96}. Therefore, Bell inequalities can be constructed in terms of the Wigner functions. This method has been utilized for testing Bell inequalities in various CV systems~\cite{BelltestCV}. Given the CM $V$ of the MRs, it is straightforward to compute the Wigner function~\cite{Barnett}. For our two-mode Gaussian state, the Wigner function is defined as the Fourier transform of the Weyl characteristic function $\chi(u)={\rm exp}(-u V u^{\rm T})$~\cite{Parisbook}, which takes the form of
\begin{equation}
W(u)=\frac{{\rm exp}(-u V^{-1} u^{\rm T})}{\pi^2 \sqrt{{\rm det} V}},
\label{WigCM}
\end{equation}
where $u$ denotes the phase-space variables associated with the quadrature fluctuation operators of $\hat u$. We then apply the displaced parity operator $\hat{\Pi}(\alpha){=}\hat{D}(\alpha)\,\hat{\Pi} \, \hat{D}^{\dagger}(\alpha)$ to be measured on each mode of the MRs, where $\hat{D}(\alpha)$ is the displacement operator $\hat{D}(\alpha){=}{\rm exp}(\alpha \hat{b}^{\dagger}{-}\alpha^* \hat{b})$ ($\alpha \, {\in} \, \mathbb{C}$) and $\hat{\Pi}$ is the parity operator, given by
\begin{equation}
\hat{\Pi}=(-1)^{\hat n}=\sum^\infty_{n{=}0}(|2n\rangle \langle 2n|-|2n+1 \rangle \langle 2n+1|),
\end{equation}
with $\hat n=\hat{b}^{\dagger} \hat{b}$ the bosonic number operator. It should be noted that in principle one could not directly make the above measurement on the MRs. However, this can be done by sending a weak red-detuned probe light, to which the state of the MR is transferred, and then measuring the probe mode~\cite{DV07}. The displaced parity measurement could be realized using a beam splitter and a photon number detector~\cite{BW96}. By using the fact that $\langle \hat{\Pi} (\alpha)\rangle{=}(\pi/2)W(\alpha)$ for each mode~\cite{BW98}, we construct the phase-space version of the Bell-Clauser-Horne-Shimony-Holt (CHSH) inequality~\cite{CHSH}, $|{\cal B}| \le 2$, with
\begin{equation}
{\cal B}=\frac{\pi^2}{4} \big[W(u_1,u_2)+W(u'_1,u_2)+W(u_1,u'_2)-W(u'_1,u'_2) \big],
\label{chsh}
\end{equation}
where $u_j{=}\{\delta q_j, \delta p_j\}$ and $u'_j{=}\{\delta q'_j, \delta p'_j\}$ ($j{=}1,2$) embody pairs of different values of the same quadrature operators of the $j$th MR. Any local realistic theory imposes the bound $|{\cal B}|\le 2$, and its violation implies that nonlocal correlation is shared by the MRs. In what follows, we define ${\cal B}_{\rm max}$ as the maximum of ${\cal B}$ optimized over the full range of $\{\delta q_1, \delta p_1, \delta q_2, \delta p_2, \delta q'_1, \delta p'_1,\delta q'_2, \delta p'_2\}$, and it is known that the maximal violation allowed by quantum mechanics is $|{\cal B}_{\rm max}|=2\sqrt{2}$~\cite{Cirelson}.

Unlike steering and entanglement, which are found in the steady state with large values, we have observed Bell violation with the parity detection only in the transient state when the mechanical modes decouple from the optical mode, which occurs at $t_m{=}2m\pi/\Delta$ ($m{=}1,2,...$) in the case of equal couplings $G_1{=}G_2$, and in the ideal parameter regime $G \gg \Delta \gg \tilde \kappa$~\cite{JieNJP,JieCF}. At these times, the mechanical entanglement can be strong. $|{\cal B}_{\rm max}|$ shows peaks at $t_m$, at which the violation of the CHSH inequality is observed with vanishing cavity decay rate $\tilde \kappa$, mechanical damping rate $\gamma$ and thermal excitations $\bar n_{1,2}$, as shown in Fig.~\ref{fig3} (a). As soon as $\gamma$ increases a little bit, $|{\cal B}_{\rm max}|$ drops rapidly and the nonlocal correlation shared by the MRs vanishes, as shown in Fig.~\ref{fig3} (b). In Fig.~\ref{fig4} we plot $|{\cal B}_{\rm max}|$ as a function of $\gamma$ and $\tilde \kappa$ at the optimal time $t=2\pi/\Delta$ for $\bar n_1\,{=}\,2\bar n_2\,{=}\,0$ and 0.05. It shows that a little rise of the thermal excitations will kill the nonlocality. This means that the Bell nonlocality of the MRs with the parity detection is extremely sensitive to any kinds of system noises. Similar finding has been observed in a hybrid atom-light-mirror system where the tripartite nonlocality is demonstrated by the violation of the Mermin-Klyshko inequality~\cite{AdP}. Since our scheme is valid with the conditions \eqref{condQLEs} fulfilled and it has been verified numerically that the scheme works optimally when $\omega_{1,2} \ge 10^2 {\rm max}\{ G_{1,2}, \tilde \kappa \}$~\cite{JieNJP}, this implies $\omega_{1,2} \ge 10^7$ Hz and mechanical Q factor $Q_m\,{=}\, \omega_{1,2}/\gamma \, {>}\,{\sim}10^7/10^{-4}=10^{11}$ for the parameters used in Fig.~\ref{fig4} in order to see the Bell violation. Taking smaller values of $G_{1,2}$, the maximum allowed values of $\gamma$ for violating the CHSH inequality also decrease keeping the Q factor $Q_m \, {>}\,{\sim} 10^{11}$. Levitated nanospheres~\cite{PNAS,Oriol,JieRC} are promising systems to achieve such a goal. Furthermore, it requires very low environmental temperature and an almost perfect cavity with extremely small cavity losses, which seems unrealistic to be implemented.

\begin{figure*}[t]
\hskip-0.6cm{\bf (a)}\hskip5.54cm{\bf (b)}\hskip5.76cm{\bf (c)}\\
\hskip-0.08cm\includegraphics[width=18.cm]{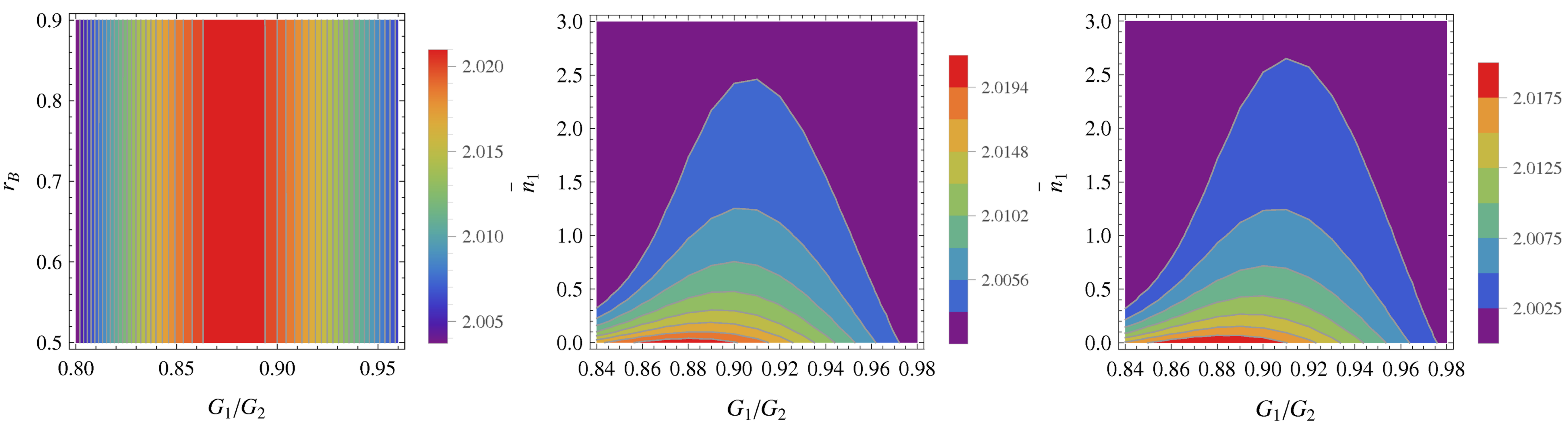}   \\
\caption{Contour plot of $|{\cal B}'_{\rm max}|$ in the steady state as a function of (a) $G_1/G_2$ and $r_B$ with $\bar n_1=\bar n_2=0$, $\gamma_1=\gamma_2=1$ Hz; (b) $G_1/G_2$ and $\bar n_1=2\bar n_2$ with $r_B=0.5$ and $\gamma_1=\gamma_2=1$ Hz; (c) $G_1/G_2$ and $\bar n_1=2\bar n_2$ with $r_B=0.5$ and $\gamma_1=\gamma_2=100$ Hz. The other parameters are $G_2=2\kappa_1=2\kappa_2=10^5$ Hz, $\Delta=0$ and $\theta=0$.}
\label{fig5}
\end{figure*}

\subsection{Bell violation with on-off detection}
\label{Bell-Q}

Although large violation of the Bell inequality has been found with dispalced parity detection, it is extremely fragile and only exists in the transient regime and in the system with extremely small noises. This is because the parity measurement detects effectively higher-order phonon number correlations and it requires very high detector efficiency or very low system noises~\cite{Lee}. Instead, the on-off detection, which measures only correlations between vacuum and phonons (or phonon absence and presence), would relax the stringent conditions for system losses in order to see the Bell violation. 

Unlike displaced parity operator, the mean value of displaced on-off detection is proportional to the Husimi Q function, $Q(\beta){=}\frac{1}{\pi}\langle \hat \Pi (\beta) \rangle$~\cite{Glauber}, with $\hat{\Pi}(\beta)$ the operator for displaced on-off detection, $\hat{\Pi}(\beta){=}\hat{D}(\beta)\,\hat{\Xi} \, \hat{D}^{\dagger}(\beta)$ ($\beta \, {\in} \, \mathbb{C}$), where $\hat{\Xi}\,{=}\,|0\rangle \langle 0 |$ represents the on-off measurement which yields an eigenvalue of $1$ for the vacuum state and $0$ for all nonzero phonon number states. In fact, $\hat{\Pi}(\beta){=}\hat{D}(\beta)\,|0\rangle \langle 0 |\, \hat{D}^{\dagger}(\beta){=}|\beta \rangle \langle \beta |$ denotes the projection onto a coherent state. In order to keep the same form of the Bell inequality $|{\cal B}'| \le 2$ for local realistic theories, we use the measurement operator $2 \hat{\Pi}(\beta) {-}\hat{\mathbb{I}}$ ($\hat{\mathbb{I}}$ is the identity operator) as the observable, which yields two possible measurement outcomes $\pm 1$. In such a way, the Bell inequality could be formulated in phase space in terms of the Q functions, analogous to the Clauser-Horne inequality~\cite{CH74}, i.e., $|{\cal B}'| \le 2$ with~\cite{Lee,BW99}
\begin{equation}
\begin{split}
{\cal B}'=&4\pi^2 \big[ Q(u_1,u_2)+Q(u'_1,u_2)+Q(u_1,u'_2)-Q(u'_1,u'_2) \big]  \\
&\,\,\,\,\,\,\,\,\, -4\pi \big[ Q(u_1)+Q(u_2) \big] +2,
\end{split}
\label{chQ}
\end{equation}
where $u_j$ and $u'_j$ ($j\,{=}\,1,2$) are defined in the same way as in Eq.\eqref{chsh}, and $Q(u_1)$ and $Q(u_2)$ are the marginal distributions of $Q(u)$. $Q(u)$ can be derived straightforwardly if the Wigner function $W(u)$ is known (by Eq.~\eqref{WigCM}) and it is a convolution of the Wigner function and a Gaussian weight~\cite{Parisbook}, i.e.,
\begin{equation}
\begin{split}
&Q(\beta_1,\beta_2)= \\
&\frac{4}{\pi^2} \!\! \int \!\!\!\! \int \!\! d^2\alpha_1 d^2\alpha_2  W(\alpha_1,\alpha_2)\, {\rm exp}\Big\{{-}2|\alpha_1{-}\beta_1|^2{-}2|\alpha_2{-}\beta_2|^2 \Big\},
\end{split}
\end{equation}
where $\beta_j=(\delta q_j+i \delta p_j)/\sqrt{2}$. $Q(\beta_1,\beta_2)$ (i.e., $Q(u)$) is therefore the Husimi Q representation of the state of the MRs. Similarly, we define ${\cal B'}_{\rm max}$ as the maximum of ${\cal B'}$ optimized over the full range of $\{\delta q_1, \delta p_1, \delta q_2, \delta p_2, \delta q'_1, \delta p'_1,\delta q'_2, \delta p'_2\}$. The violation of $|{\cal B}'_{\rm max}| \le 2$ implies the presence of nonlocal correlations shared by the MRs.

In Fig.~\ref{fig5} we show $|{\cal B'}_{\rm max}|$ in the steady state as a function of some key parameters of the system. As expected, the Bell violation with the on-off detection is not sensitive to cavity losses and $|{\cal B}'_{\rm max}|$ is {\it almost} unchanged by altering the effective cavity decay rate (realized by adjusting $r_B$), as shown in Fig.~\ref{fig5} (a). $|{\cal B}'_{\rm max}| \le 2$ is violated even with large values of the cavity decay rate. This overcomes the biggest obstacle for observing the Bell violation with the parity measurement. However, since it is in the steady state, which is more affected by various noises than in the transient state, the violation of the Bell inequality is only moderate. One may expect larger violation occurs in the transient state in the case of equal couplings $G_1=G_2$. However, after a careful check, we find only tiny Bell violation in the case of $G_1=G_2$. This may be due to the fact that $G_1=G_2$ is not optimal for Bell violation with the on-off detection (the optimal values of $G_1/G_2$ are away from 1, see Fig.~\ref{fig5}). Fig.~\ref{fig5} (b) and (c) show that $|{\cal B'}_{\rm max}|$ is sensitive to the thermal excitations $\bar n_{1,2}$ but not so sensitive to the mechanical damping rate $\gamma$: $|{\cal B'}_{\rm max}|$ drops only a bit when $\gamma$ increases from 1 Hz to 100 Hz. In order to see Bell violation, the system must be at very low environmental temperature, e.g., for nanogram-sized MRs of frequencies ${\sim} 10^8$ Hz~\cite{OMRMP}, $\bar n_{1,2}\, {\sim} \, 1$ implying that the temperature must be as low as ${\sim} \, 1$ mK, which is still quite challenging. For more massive MRs with typically lower characteristic frequencies~\cite{Connell}, Bell violation requires even lower temperature which poses a greater challenge to the experiment.

\section{conclusions and remark}
\label{concl}

We have studied nonlocal properties, specifically EPR steering and Bell nonlocality, of two macroscopic MRs in optomechanical systems. We have shown that large and robust one-way and two-way steerings could be achieved in the steady state with realistic parameters, and analyzed the mechanism accounting for the asymmetric nature of steering. Furthermore, we have tested Bell inequalities in phase space based on displaced parity and on-off measurements, respectively. For displaced parity detection, large Bell violation is observed in the transient state but it requires extremely small system noises and dissipation rates. In contrast, for displaced on-off detection, moderate Bell violation is found in the steady state and the nonlocality is robust against cavity losses and mechanical damping. For both the measurements, very low environmental temperature is required in order to violate Bell inequalities, which is the main obstacle for the case with on-off detection. Our work offers a possible answer in the framework of standard quantum mechanics for the lack of observations of quantum correlations shared by macroscopic/massive objects. 

We remark that throughout the paper we have assumed perfect state transfer from MRs to probe modes and perfect realization of the displaced parity/on-off detection (Gaussian measurements for steering), and have not considered any technical imperfections, such as detector inefficiencies, dark counts, various technical noises, and so on. A serious proposal for an actual experimental test should include all of these effects. Here for simplicity we have neglected these imperfections.

\section*{ACKNOWLEDGMENT}

J.L. thanks G. Adesso for discussions on Gaussian steering, and S. Zippilli and D. Vitali for their careful review and feedback on this work. This work has been supported by the Joint Fund of the National Natural Science Foundation of China (Grant No. U1330203).


\end{document}